# Effects of 3 MeV Proton Irradiation on Superconductivity and CDW in 2$H$-NbSe$_2$ Single Crystals

Wenjie Li[*], Sunseng Pyon, Akiyoshi Yagi, Tong Ren, Masahiro Suyama, Jiachen Wang, Takumi Matsumae, Yuto Kobayashi, Ayumu Takahashi, Daisuke Miyawaki, and Tsuyoshi Tamegai

*Department of Applied Physics, The University of Tokyo, 7-3-1 Hongo, Bunkyo-ku, Tokyo 113-8656, Japan*

Interplay between superconductivity and charge-density wave (CDW) in 2$H$-NbSe$_2$ single crystals irradiated by 3 MeV protons is studied. Both $T_c$ and $T_{CDW}$ are found to decrease monotonically with the increase in irradiation dose. This behavior is different from electron-irradiated NbSe$_2$, where $T_{CDW}$ is suppressed monotonically with the increase in dose, while $T_c$ shows an initial enhancement before it starts to decrease. We attempt to explain this difference based on the *negative pressure* effect which has been reported in our previous study on NbSe$_2$ irradiated by heavy ions.

## 1. Introduction

Artificial defects introduced by particle irradiations can act as pair-breakers and suppress $T_c$ of superconductors [1-5]. However, the irradiation can also enhance $T_c$ in superconductors with competing ground states (such as charge-density wave, CDW) [6-7]. One explanation for such $T_c$ enhancement is based on the fact that superconductivity and CDW occur by using different parts of the Fermi surface. After the particle irradiation, CDW is suppressed and parts of the gapped Fermi surface will be released, which can be used to enhance the superconductivity. The competition between superconductivity and CDW has been observed in high-temperature cuprate superconductors [8-12], Kagome superconductor CsV$_3$Sb$_5$ [13-16], and low-temperature superconductors such as Lu$_5$Ir$_4$Si$_{10}$ [7] and NbSe$_2$ [6, 17-19]. Modifications of electronic states either by chemical doping or by physical pressure are typical methods which are usually used to study the relationship between superconductivity and CDW. Recently, the third method has been applied to the study of competition between superconductivity and CDW in terms of irradiation [6]. However, the results obtained by these methods are different in NbSe$_2$. For the experiments using hydrostatic pressure [17-18] and electron irradiation [6] on NbSe$_2$ single crystals, with increasing pressure or irradiation dose, $T_c$ shows an initial enhancement, and after $T_c$ reaching a maximum value it starts to decrease monotonically. In the case of hydrostatic pressure experiments, $T_c$ can be enhanced from 7.2 K up to ~8.5 K [17-18]. In the case of 2.5 MeV electron irradiation, $T_c$ has been enhanced from 7.25 K to ~7.45 K [6], followed by monotonic suppression by further increase in irradiation dose. On the other hand, through Te substitution into Se site in NbSe$_2$, $T_c$ is monotonically decreased along with a monotonic enhancement of $T_{CDW}$ [19]. All these results make it difficult to claim how superconductivity and CDW influence to each other in NbSe$_2$. So it is necessary to study the interplay between superconductivity and CDW in NbSe$_2$ in more detail.

In this paper, the interplay between superconductivity and CDW in NbSe$_2$ is studied by introducing artificial defects through 3 MeV proton irradiation. We followed the evolution of temperature dependence of resistivity via *in situ* resistivity measurements between successive irradiations, which are followed by the inspection of the crystal lattice with increasing dose.

## 2. Experimental Details

Single crystals of 2*H*-NbSe$_2$ were prepared by iodine vapor transport method as described in Ref. [20]. 3 MeV proton irradiation experiments were conducted at NIRS-HIMAC in Chiba, Japan. Before the irradiation, single crystals were prepared into thin plates with thickness of ~20 $\mu$m, which is thinner than the projected range of 3 MeV protons in NbSe$_2$ (~53 $\mu$m). The projected range is calculated by the stopping and range of ions in matter (SRIM-2008) [21]. For *in situ* resistivity measurements, gold wires with diameter of 25 $\mu$m were attached on the surface of the sample by silver paste in a standard four-probe configuration. After the sample was loaded onto a sapphire plate, the single crystal was cooled down by a closed-cycle refrigerator at the end of the irradiation port. The surface without gold wires was irradiated by protons. Irradiation was performed at 40 K to avoid annealing of created defects. The in-plane resistivity was measured by using both AC resistance bridge (LR-700, Linear Research) and DC nanovoltmeter (2182A, Keithley) with excitation current of 2 mA. The crystal structure was characterized at room temperature by using a commercial diffractometer (Smartlab, Rigaku) with Cu *K*α radiation. After determining diffraction angles, Bragg's law was used to calculate corresponding *d*-spacings, which are related to lattice parameters *a* and *c* by the relation for hexagonal lattice of

$$\frac{1}{d^2} = \frac{4(h^2 + hk + k^2)}{3a^2} + \frac{l^2}{c^2}, \tag{1}$$

where *h*, *k*, and *l* are Miller indices. In general, the Bragg peaks other than (00*l*) peaks cannot be directly obtained by using standard $\omega$-$2\theta$ scan for single crystals ($\omega$ is the sample angle and $2\theta$ is the detector angle). According to the Laue condition for the constructive interference of

$$\Delta \mathbf{k} = \mathbf{k}_{out} - \mathbf{k}_{in} = \mathbf{G}, \quad (2)$$

where $\mathbf{k}_{out}$, $\mathbf{k}_{in}$, and $\mathbf{G}$ are outgoing wave vector, incoming wave vector, and reciprocal lattice vector, respectively. ($h0l$) reciprocal lattice points can be brought onto the Ewald sphere by rotating the sample. Through successive optimizations of $\omega$ and $2\theta$, we can determine $2\theta$ for ($10\bar{1}0$) peaks, from which we can determine $d$-spacing for ($10\bar{1}0$) plane. The $c$ lattice parameter is calculated by taking an average value from (004), (006), and (008) peaks. For the calculation of $a$ lattice parameter, the average value was taken from the observed six equivalent ($10\bar{1}0$) peaks. The magnetization measurements were conducted by using a commercial SQUID magnetometer (MPMS-XL5, Quantum Design).

## 3. Results and Discussion

### 3.1 In situ resistivity measurements

Based on the previous studies on the CDW in $NbSe_2$ [22-23], the CDW transition in $NbSe_2$ single crystal is relatively weak and the sample quality has strong influence on the formation of CDW. Thus, it is important to choose high-quality samples with low residual resistivity for the experiments. For that purpose, pristine samples used here have relatively large residual resistivity ratios RRR $\equiv \rho(300\ K)/\rho(8\ K) \sim 50$. The CDW transition can be clearly observed at $T = 33\ K$ in the pristine sample as shown in Fig. 1(a). With increasing proton dose, resistivity increases monotonically due to the effect of introduced artificial defects. Figure 1(b) is the enlarged temperature dependence of resistivity with increasing dose near $T_c$, where monotonic suppression of $T_c$ can be clearly observed. To clearly identify the CDW transition, the temperature derivative of $\rho$, $d\rho/dT$, around the CDW transition is calculated as shown in Fig. 1(c). We define $T_{CDW}$ as the local minimum of $d\rho/dT - T$ curves marked by arrows in Fig. 1(c). It is clear that $T_{CDW}$ decreases monotonically with the increase in irradiation dose and it cannot be defined at the irradiation dose of $6\times10^{16}/cm^2$ or more. Figures 1(d) and (e) show evolution of $T_{CDW}$ and $T_c$ with increasing dose. Here, to make the quantitative comparison with electron irradiation [6] easy, the dose is replaced by the change in resistivity at $T = 40\ K$, $\Delta\rho$ (40 K). When the CDW is formed, the resistivity value below $T_{CDW}$ should be enhanced compared with the putative value without CDW. This feature can provide an alternative way to identify the presence of CDW. In other words, difference in resistivity changes due to irradiation at the temperature above $T_{CDW}$ and just above $T_c$ can be a measure of CDW order. Actually, as shown in Fig. 1(f), $\Delta\rho$ (8 K) - $\Delta\rho$ (40 K) changes monotonically as CDW is suppressed by proton irradiation, and it saturates at a disorder level with $\Delta\rho$ (40 K) $\sim$10 $\mu\Omega\cdot cm$, above which CDW is not observed any more as seen in Fig. 1(d). Very similar behavior of $\Delta\rho$ (7.6 K) – $\Delta\rho$ (40 K) has also been reported in electron-irradiated NbSe$_2$ [6]. However, it should be noted that the increasing trend for $\Delta\rho$ (8 K) – $\Delta\rho$ (40 K) with disorder is opposite to what is expected for a system with CDW. Let us consider a system with two metallic bands, band 1 forming CDW ($\rho_1$: red line) and band 2 responsible for superconductivity ($\rho_2$: blue line) as shown in the inset of Fig. 1(f). For simplicity, we assume that band 1 turns into an insulator with infinite resistivity below $T_{CDW}$ and the total resistivity of the system can be approximated by a parallel circuit of the two bands. In such a case, total resistivity should behave as shown by the black line ($\rho_0$) in the inset of Fig. 1(f). After complete suppression of CDW by irradiation-induced disorder, both bands should show metallic conduction with enhanced resistivity due to disorder ($\rho_1'$: broken red line and $\rho_2'$: broken blue line). Hence, the total resistivity after the suppression of CDW ($\rho_0'$) should behave as shown by the broken black line in the inset of Fig. 1(f). Now it is clear that $\Delta\rho_a \equiv \Delta\rho$ (8 K) is smaller than $\Delta\rho_a \equiv \Delta\rho$ (40 K), making $\Delta\rho$ (8 K) – $\Delta\rho$ (40K) negative. Obviously, this decreasing trend of $\Delta\rho$ (8 K) – $\Delta\rho$ (40 K) with disorder is opposite to what we have observed experimentally. Such counter-intuitive behavior of resistivity at low temperatures with disorder in a system with CDW requires further detailed studies.

According to the Anderson theorem, nonmagnetic disorder will not affect $T_c$ if a superconductor has an isotropic $s$-wave gap structure [24]. For NbSe$_2$, the anisotropic $s$-wave gap has been reported by thermal transport [25], angle-resolved photoemission spectroscopy [26], and scanning tunneling microscope measurements [27]. This means that the nonmagnetic disorder can act as pair-breaker and affect the $T_c$ of NbSe$_2$. To discuss the pair-breaking effect due to nonmagnetic scattering quantitatively, we estimated the normalized scattering rate ($g$). $g$ can be calculated based on the Drude model,

$$g = \frac{\hbar}{2\pi k_B T_{c0} \tau}, \quad (3)$$

where $\hbar$, $k_B$, and $\tau$ are the Planck's constant (divided by $2\pi$), the Boltzmann constant, and the scattering time, respectively. In multi-gap superconductors, the scattering time includes at least two components, namely the interband scattering time $\tau_{inter}$ and intraband scattering time $\tau_{intra}$. Both $\tau_{inter}$ and $\tau_{intra}$ play important roles to affect $T_c$ and resistivity [28-29]. In the simplest case, if these two scattering times are the same, $\tau_{inter} = \tau_{intra} = \tau$,

$$\frac{1}{\Delta\rho_0} = \frac{1}{\Delta\rho_{intra}} + \frac{1}{\Delta\rho_{inter}} = \left(\frac{ne^2}{m^*}\right)(\tau_{intra} + \tau_{inter}) = \frac{2ne^2}{m^*\tau}. \quad (4)$$

By inserting Eq. (4) into Eq. (3), the normalized scattering rate can be calculated as

$$g = \frac{\hbar ne^2 \Delta\rho_0}{\pi k_B T_{c0} m^*}, \quad (5)$$

where $n$ is the carrier density, $\Delta\rho_0 = \rho_0^i - \rho_0^0$ (the superscript $i$ represents the $i$-th irradiation and the zero-temperature resistivity ($\rho_0$) was extrapolated through fitting the resistivity above

$T_c$ to the function of $\rho = \rho_0 + aT^2$ (the dashed line in Fig. 1(a)), and $m^*$ is the effective mass of the quasiparticle. Eq. (5) is only applicable to superconductors with multiple gap and equal scattering times. In the case of NbSe$_2$, its gap structure is still under debate. Evidences supporting two-gap feature have been reported in specific heat study [30], penetration depth measurement [31], pressure measurement [32], scanning tunneling spectroscopy [33], and quantum dot-assisted spectroscopy measurements [34]. There are also some experimental observations of single-gap with an anisotropic Fermi surface in NbSe$_2$ [35-36]. These debates make the accurate calculation of scattering rate $g$ for NbSe$_2$ complicated. However, the purpose of our calculation of $g$ is to compare the data on the effect of proton irradiation with the electron irradiation. So, for the sake of reasonable comparison, we choose the same treatment on $g$ as in Ref. [37], where the interband scattering rate is considered to be 0. Thus, a factor of 1/2 should be multiplied in Eq. (4).

To simplify the calculation of $g$, the penetration depth $\lambda_0 = (m^*/\mu_0 n e^2)^{1/2}$ is used, which can allow us to avoid direct estimation of $n$ and $m^*$. Then the formula for $g$, which is called $g^\lambda$, becomes

$$g^\lambda = \frac{\hbar \Delta \rho_0}{2\pi k_B T_{c0} \mu_0 \lambda^2}. \tag{6}$$

By using the penetration depth ($\lambda = 1250$ Å) of NbSe$_2$ reported in Ref. [31], the normalized $T_c$ ($t_c = T_c/T_{c0}$) dependence of $g^\lambda$ is plotted in Fig. 2. In Fig. 2, the curve for the suppression of $T_c$ described by the Abrikosov-Gork'ov (AG) theory for magnetic pair-breaking is also included. AG theory can be described as

$$\ln(t_c) = \psi\left(\frac{1}{2} + \frac{g}{2t_c}\right) - \psi\left(\frac{1}{2}\right), \tag{7}$$

where $\psi$ is the digamma function. As a reference, the data for NbSe$_2$ after 2.5 MeV electron irradiation [37] are also added. Except for the small $g^\lambda$ range with complicated behavior in the case of electron irradiation, $T_c$ suppression rate in NbSe$_2$ after proton irradiation (~0.03 K/$\mu\Omega$·cm) is smaller than that in electron irradiated crystals (~0.05 K/$\mu\Omega$·cm [37]). This indicates that the defects introduced by proton irradiation act as weaker pair-breakers. One explanation for this difference is that the defects introduced by proton irradiation have larger dimension (cascade of point defects) than the defects introduced by electron irradiation (defects introduced by electron irradiation are mainly Frenkel pairs), the larger defects induce scattering with smaller wave numbers in reciprocal space, leading to weaker pair breaking.

Similar phenomena have also been observed in other superconductors. For example, by irradiating Ba$_{1-x}$K$_x$Fe$_2$As$_2$ single crystals with 2.5 MeV electrons, $T_c$ suppression rate is ~0.2

K/$\mu\Omega$·cm [5]. On the other hand, for Ba$_{1-x}$K$_x$Fe$_2$As$_2$ irradiated by 3 MeV protons, $T_c$ suppression rate is weaker with suppression rate ~0.1 K/$\mu\Omega$·cm [4]. In V$_3$Si single crystals, $T_c$ suppression rate is 0.013 K/$\mu\Omega$·cm by 2.5 MeV electron irradiation [37], while it is 0.006 K/$\mu\Omega$·cm by 35 MeV proton irradiation [38]. These data suggested that defects introduced by electron irradiation act as stronger pair-breakers and affect $T_c$ more than defects introduced by proton irradiation.

*3.2 Expansion of lattice parameters*

Materials in the ground state are expected in the state with the maximum density with minimum lattice parameters. It means that whenever defects are introduced by energetic particle irradiation, the crystal lattice is expected to expand to some extent. Actually, when materials are irradiated by heavy-ions, during the process of rapid cooling of the melted crystal amorphous tracks are created, which are known to have lower density with larger separation of constituent atoms. Due to the expansion of these amorphous regions, the surrounding lattice is forced to expand. Such lattice expansion has been observed in many materials after heavy-ion irradiations [20, 39-44]. In some superconductors, for example, the $c$-axis lattice parameter expansions have been observed at rates of $dT_c/dB_\Phi$ = 0.021%/T (3.8 GeV Ta irradiation), 0.050%/T (80 MeV I irradiation), 0.081%/T (200 MeV I irradiation), and 0.163%/T (120 MeV Au irradiation) on ErBa$_2$Cu$_3$O$_7$ thin films [43], where $B_\Phi$ is the dose equivalent matching field ($B_\Phi$ = 1 T corresponds to 5 × 10$^{10}$/cm$^{-2}$ defects). The lattice expansion has also been observed in our previous study on NbSe$_2$ single crystals after introducing columnar defects by heavy-ion irradiations, where the $c$-axis lattice parameters expanded at rates of 0.016%/T (800 MeV Xe) and 0.030%/T (320 MeV Au) [44]. Similarly, the lattice parameter expansion is also observed in NbSe$_2$ after 3 MeV proton irradiation in the present study.

Figures 3(a) and (b) show evolutions of (004) and (10$\overline{1}$0) Bragg peaks with increasing irradiation dose for NbSe$_2$ single crystals, where Bragg peaks shift monotonically from higher angle to lower angle. The full width at half maximum (FWHM) in Figs. 3(c) and (d) show almost no change after 3 MeV proton irradiation, which indicates that the lattice expands uniformly by 3 MeV proton irradiation. As shown in Figs. 3(e) and (f), both $c$ and $a$ increase linearly at rates of 0.012%/unit dose and 0.011%/unit dose (unit dose equals 1×10$^{16}$/cm$^2$), respectively. Lattice parameters $a$ and $c$ are expanded at a similar rate after 3 MeV proton irradiation, which is different from the case of heavy-ion irradiations [44]. For NbSe$_2$ introduced with columnar defects by 320 MeV Au irradiation, expansion rate of the *c*-axis ($dc/dB_\Phi$ ~0.030%/T) is about twice larger than that of the *a*-axis ($da/dB_\Phi$ ~0.016%/T), at the same time, $T_c$ was found to be suppressed almost linearly at a rate of 0.07 K/T [44]. It should be noted that the mechanism of lattice expansion by proton and heavy-ion is different. In the case of proton irradiation, the lattice change is mainly coming from the collision cascades between proton and target atom. In other words, the probability of a proton hitting the target atom is the same in different directions, which is consistent with the above results reporting a similar expansion rate for both *a*-axis and *c*-axis. On the other hand, heavy-ion irradiation creates separated linear amorphous tracks to expand lattice parameters, the effect for different direction should not be the same. In previous studies on $Nb_3Sn$ irradiated by protons and heavier He ions, lattice parameter of the sample irradiated by He expanded more under the same irradiation dose; 0.005% [45] and 0.047% [46] at a dose of $1 \times 10^{16}$/cm$^2$ for proton and He, respectively. These results also suggest that the lattice expansion induced by irradiations of heavier ions is stronger.

*3.3 $T_c$ suppression after 3 MeV proton irradiation*

Based on the above results, it is plausible that 3 MeV proton irradiation induces lattice expansion in $NbSe_2$ together with the monotonic suppression of $T_c$. In the study of hydrostatic pressure on $NbSe_2$ single crystals, $T_c$ is enhanced with the shrinkage of lattice [47]. This means that $T_c$ suppression and lattice expansion in $NbSe_2$ after 3 MeV proton irradiation can be considered as the effect of *negative pressure*. Figure 4(a) shows the dose dependence of $T_c$, which is estimated from the magnetization measurements as shown in the inset of Fig. 4(a). Monotonic decrease in $T_c$ with increasing dose can be observed.

To accurately calculate how much $T_c$ is suppressed by lattice expansion, it is necessary to separate the effect related to change in *a*-axis and in *c*-axis. For that purpose, the data for uniaxial pressure derivative of $T_c$ along *a*-axis ($dT_c/dp_a$) and *c*-axis ($dT_c/dp_c$) are necessary. The hydrostatic pressure and the uniaxial pressure derivative of $T_c$ are related by the formula of

$$\frac{dT_c}{dp} = 2\frac{dT_c}{dp_a} + \frac{dT_c}{dp_c}. \qquad (8)$$

By using reported data on $dT_c/dp_c$ [48] and $dT_c/dp$ [47] for $NbSe_2$ single crystals, $dT_c/dp_a$ can be indirectly evaluated as $dT_c/dp_a = 0.10$ K/kbar. By combining the linear compressibilities ($\Delta a/a/\Delta p = 4.1\ (\pm 0.4) \times 10^{-4}$ /kbar and $\Delta c/c/\Delta p = 16.2\ (\pm 0.5) \times 10^{-4}$ /kbar [47]) of $NbSe_2$ with the uniaxial pressure derivative of $T_c$ for both *a*-axis and *c*-axis, the *a*-axis and *c*-axis lattice variation induced $T_c$ changes are calculated as $dT_c/(\Delta a/a)$ = -244 K and $dT_c/(\Delta c/c)$ = 91 K. By connecting these data with the lattice expansion data for NbSe$_2$ irradiated by 3 MeV proton, $T_c$ is expected to change -0.18 K by *a*-axis expansion and 0.07 K by *c*-axis expansion at a dose of $7\times10^{16}$/cm$^2$, resulting in a total $T_c$ change of -0.29 K at this dose. This value is about ~46% of the experimental value of -0.63 K. As we mentioned above, factors which are known to affect $T_c$ of NbSe$_2$ besides lattice expansion are disorder and CDW. For the effect of CDW on $T_c$ in NbSe$_2$, how it quantitatively affects $T_c$ is not yet clear. There are contradictory reports claiming enhancement [6, 17-18, 49] and suppression [50] of $T_c$ by the destruction of CDW. $T_c$ enhancement by the destruction of CDW can be understood based on the competition between superconductivity and CDW. On the other hand, $T_c$ suppression by the destruction of CDW is explained based on the hardening of phonons leading to weaker electron-phonon interaction. According to the electron irradiation experiment on NbSe$_2$ [6], after CDW is completely destroyed by irradiation, $T_c$ decreases almost linearly with the increase in irradiation dose. Based on the suppression rate of $T_c$ at high dose ($dT_c/d\rho$ ~0.05 K/$\mu\Omega\cdot$cm) and assuming that this rate is independent of dose, we can estimate the $T_c$ suppression induced by the corresponding disorder at the dose which forms the maximum $T_c$, and calculate how much $T_c$ can be enhanced by the destruction of CDW as |$\Delta T$ (CDW)| = |$\Delta T$ (disorder)| + |$\Delta T$ (measured value)|. The $\Delta T$ (CDW) at the maximum $T_c$ in Ref. [6] is calculated to be 0.48 K (|$\Delta T$ (disorder)| = |$\Delta\rho$| × |$dT_c/d\rho$| = 0.28 K, |$\Delta T$ (measured value)| = 0.2 K). We assume that this value also applies to the case of 3 MeV proton irradiation. With this assumption, the disorder-induced $T_c$ suppression in NbSe$_2$ after 3 MeV proton irradiation can be estimated as |$\Delta T_c$ (disorder)| = |$\Delta T_c$ (measured value)| – |$\Delta T_c$ (lattice expansion)| + |$\Delta T_c$ (CDW)|. As an example, the disorder-induced $T_c$ suppression for sample irradiated by 3 MeV proton irradiation with the dose of $7\times10^{16}$/cm$^2$ is calculated to be 0.63 K - 0.29 K + 0.48 K = 0.82 K. This result indicates that the $T_c$ suppression induced by disorder is stronger than that of lattice expansion, which is different from the effect of columnar defects created by 320 MeV Au irradiation. In the case of 320 MeV Au irradiated NbSe$_2$ single crystals, the $T_c$ suppression induced by lattice expansion is stronger than that by disorder [44].

Based on the above results, we may understand why the initial $T_c$ enhancement is not observed in NbSe$_2$ single crystals irradiated by 3 MeV proton irradiation. Effects of proton irradiation on $T_c$ can be divided into three factors, which are disorder, lattice expansion, and suppression of CDW as shown schematically in Fig. 4(b). At small proton irradiation doses, CDW is suppressed by disorder and the gapped Fermi surface is recovered. When CDW is mostly suppressed at a critical dose $\Phi^*$, $T_c$ enhancement through this channel should stay constant above $\Phi^*$. As a first approximation, we can assume that the change in $T_c$ through this channel is a linear function of the dose below $\Phi^*$. With the combination of all these three factors, $T_c$ is expected to decrease monotonically as shown by the broken line in Fig. 4(b). As for the initial $T_c$ enhancement observed in the electron irradiation experiments [6], there are two possible factors that may help us to understand it. The first is that the change in lattice parameters in electron irradiation for NbSe$_2$ may be ignored, since compared with the cascade point defects introduced by proton irradiation, the Frenkel pairs introduced by electron irradiation have a weaker effect on lattice parameters. Actually, we have performed XRD experiments on electron irradiated samples (some of 2.5 MeV electron irradiated NbSe$_2$ single crystals used in Ref. [6]) to observe the change in lattice parameters. In these measurements, no prominent lattice variations have been observed in samples irradiated with doses of 1.0 C and 1.6 C. Unfortunately, however, we cannot conclude that negligible lattice variation occurs in electron irradiated crystals, since the $T_c$ of these crystals have returned almost to the pristine value due to long-term room temperature annealing. Another factor is the effect of CDW on $T_c$. As mentioned above, the effect of disorder on $T_c$ in electron-irradiated sample is stronger than that in proton-irradiated samples. So if only the effects from CDW and disorder on $T_c$ for electron irradiation are considered, the initial $T_c$ enhancement should not be observed in NbSe$_2$ irradiated by electron. However, by comparing the dose dependence of $T_{CDW}$ for electron irradiation and proton irradiation, we found that $T_{CDW}$ is suppressed faster in electron irradiation [6] than proton irradiation. This fact may suggest that $T_c$ enhancement by CDW is faster in the electron irradiated samples below $\Phi^*$. If this is true, it is possible to observe an initial $T_c$ enhancement by considering the effects of disorder and CDW in small electron irradiation dose. It is interesting to note that $T_c$ in Nb$_{1-x}$Ta$_x$Se$_2$ with atomic scale disorder also show nonlinear initial enhancement of $T_c$ [49].

4. Summary

We conducted *in situ* resistivity measurements on 2*H*-NbSe$_2$ single crystals irradiated by 3 MeV protons. Unlike the effects of 2.5 MeV electron irradiation on NbSe$_2$ single crystals, introduction of low-density point defects did not enhance superconductivity. Instead, both $T_{CDW}$ and $T_c$ were monotonically suppressed with the increase in irradiation dose. Weak but finite lattice expansions along both *a*-axis and *c*-axis were observed after the 3 MeV proton irradiation. By analyzing the lattice-expansion-induced $T_c$ suppression based on the *negative pressure* effect with that based on disorder due to proton irradiation on NbSe$_2$, $T_c$ was found to be suppressed more by disorder rather than by lattice expansion. This is different from the effect of columnar defects on NbSe$_2$ single crystals. In addition, comparison of the $T_c$ suppression rate after proton and electron irradiations, it turns out that defects introduced by proton irradiation in NbSe$_2$ act as weaker pair-breakers, which is consistent with the case for Ba$_{1-x}$K$_x$Fe$_2$As$_2$ and V$_3$Si single crystals.

**Acknowledgment**

We would like to thank Prof. R. Prozorov for providing us the 2.5 MeV electron irradiated NbSe$_2$ single crystals for XRD experiments as well as illuminating discussions.

*E-mail: wenjiecd@gmail.com

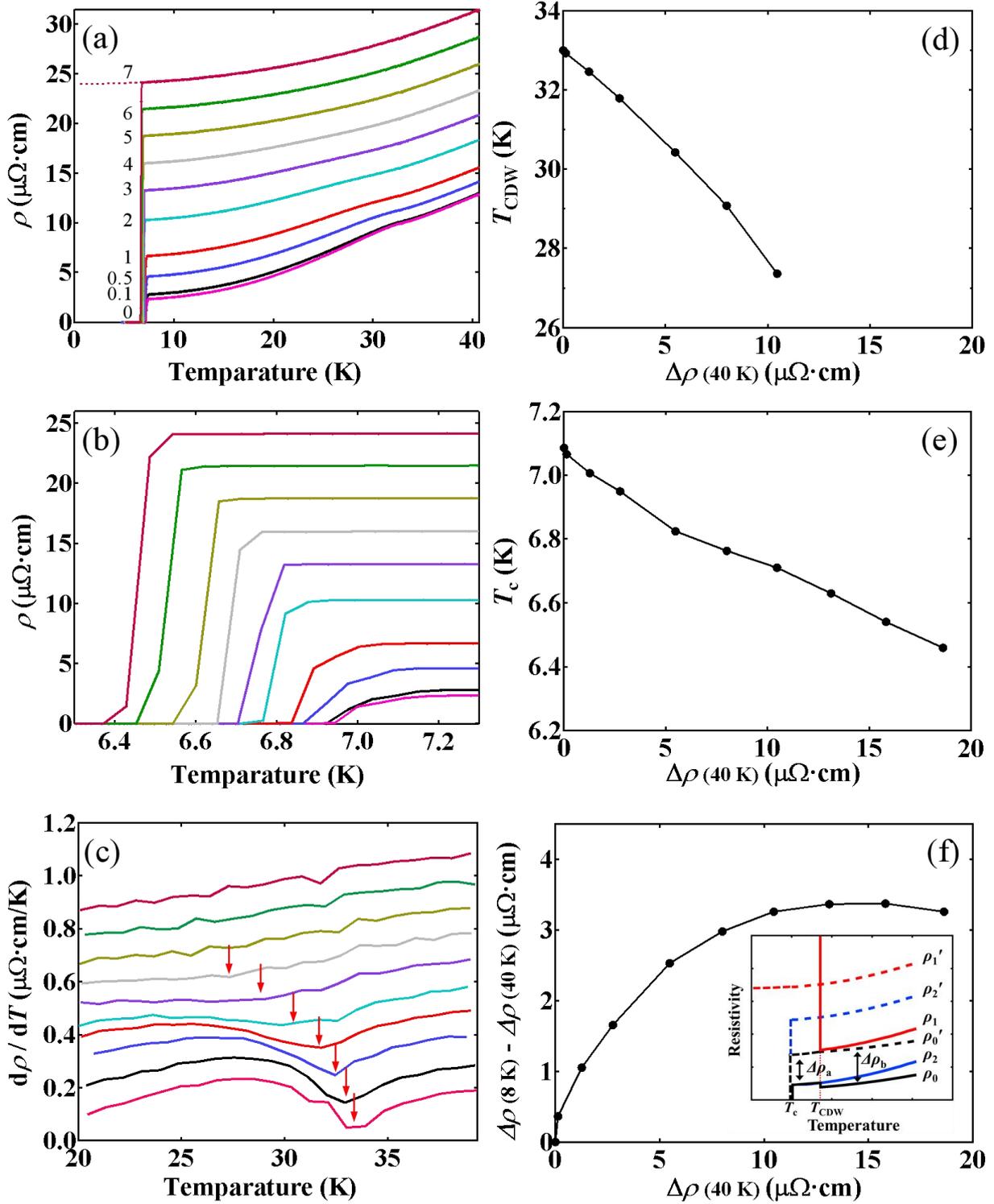

Fig. 1. (Color online) (a) *In situ* resistivity measurement results for NbSe$_2$ single crystal irradiated by 3 MeV protons up to a maximum dose of 7×10$^{16}$/cm$^2$. (b) The enlarged $\rho$ -$T$ curves around $T_c$ region. (c) The derivative of temperature dependence of resistivity. Disorder (evaluated by $\Delta\rho$ at 40 K) dependence of (d) $T_{CDW}$, (e) $T_c$, and (f) $\Delta\rho$ (8 K) – $\Delta\rho$ (40 K). The inset of (f) shows a schematic diagram of temperature dependence of resistivity for a system with two bands, band 1 forming CDW and band 2 responsible for superconductivity before and after the introduction of defects. Refer to the main text for definitions of all labels.

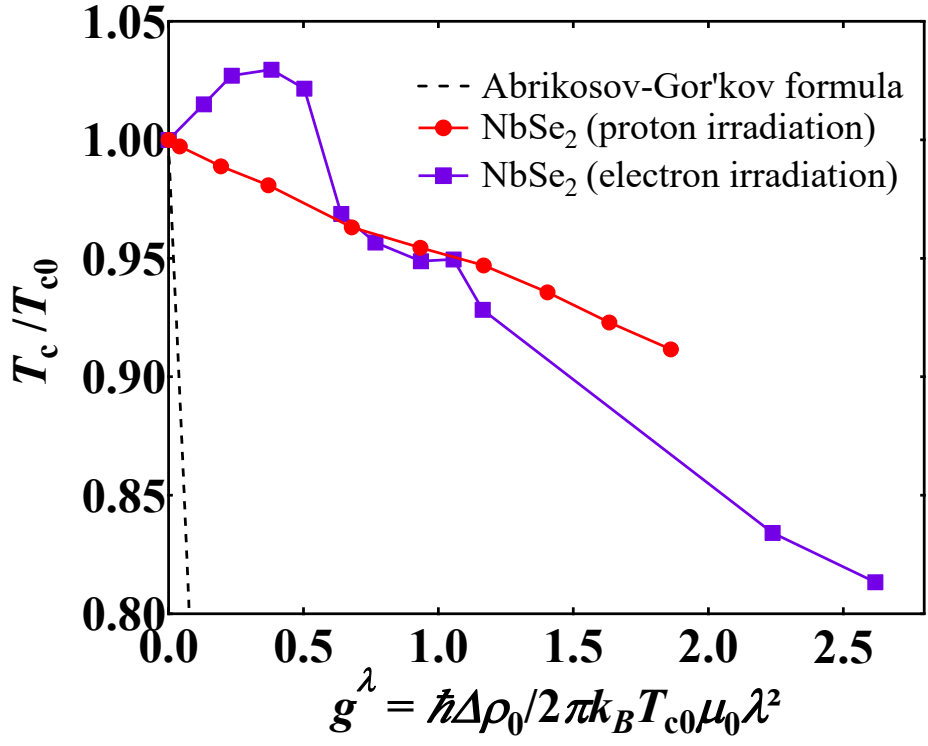

Fig. 2. (Color online) $T_c/T_{c0}$ as a function of a normalized scattering rate in NbSe$_2$ single crystal irradiated by 3 MeV protons evaluated by London penetration depth $g^\lambda = \hbar\Delta\rho_0/(2\pi k_B T_{c0}\mu_0\lambda_0^2)$. Data for NbSe$_2$ single crystals irradiated by 2.5 MeV electrons is also included [37].

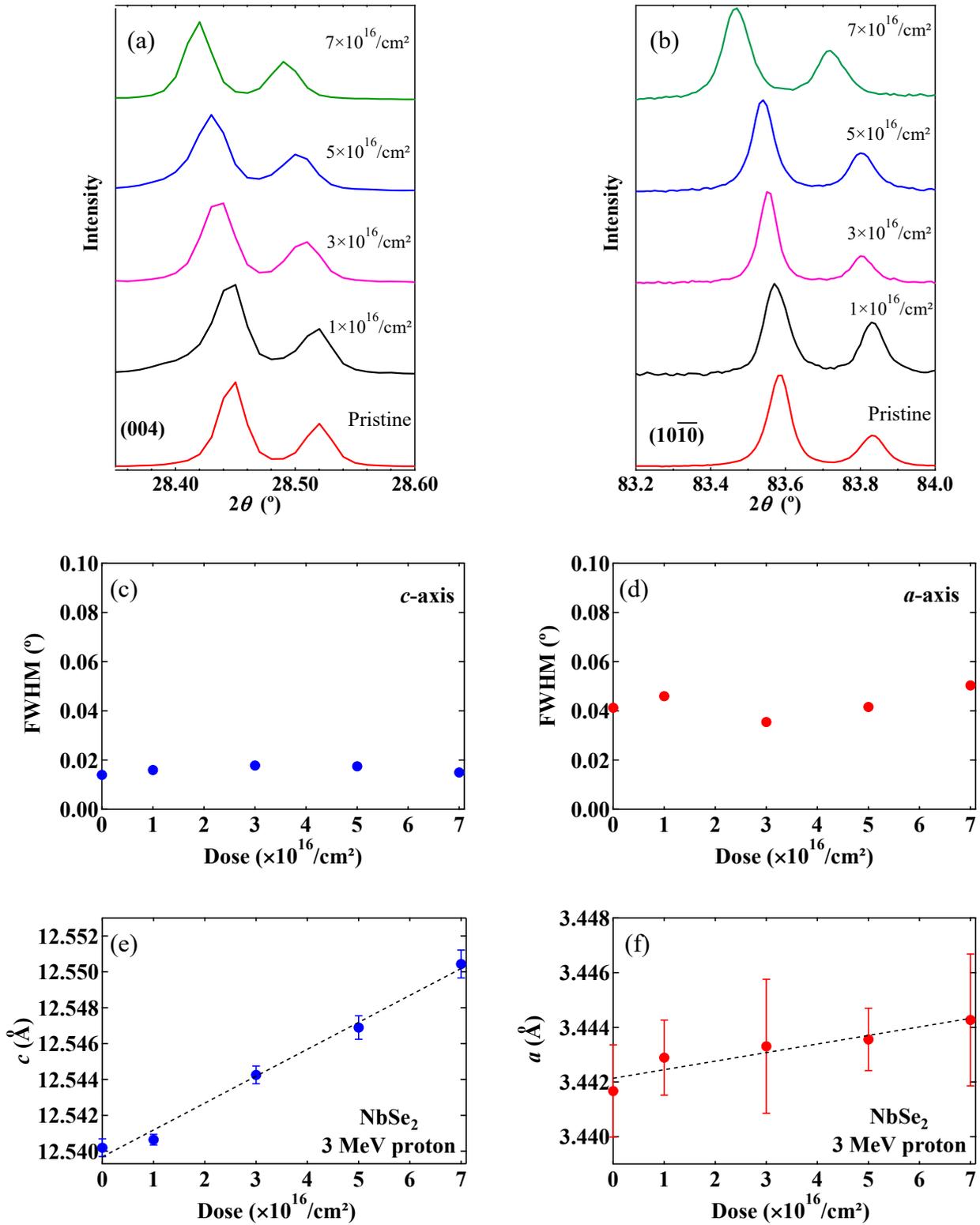

Fig. 3. (Color online) (a) (004) and (b) (10$\bar{1}$0) diffraction peak profiles for NbSe$_2$ single crystals before and after 3 MeV proton irradiation. Dose dependences of FWHM of (c) (004) and (d) (10$\bar{1}$0) peaks for NbSe$_2$ single crystals after 3 MeV proton irradiation. Dose dependences of (e) $c$-axis and (f) $a$-axis lattice parameters for NbSe$_2$ single crystals after 3 MeV proton irradiation.

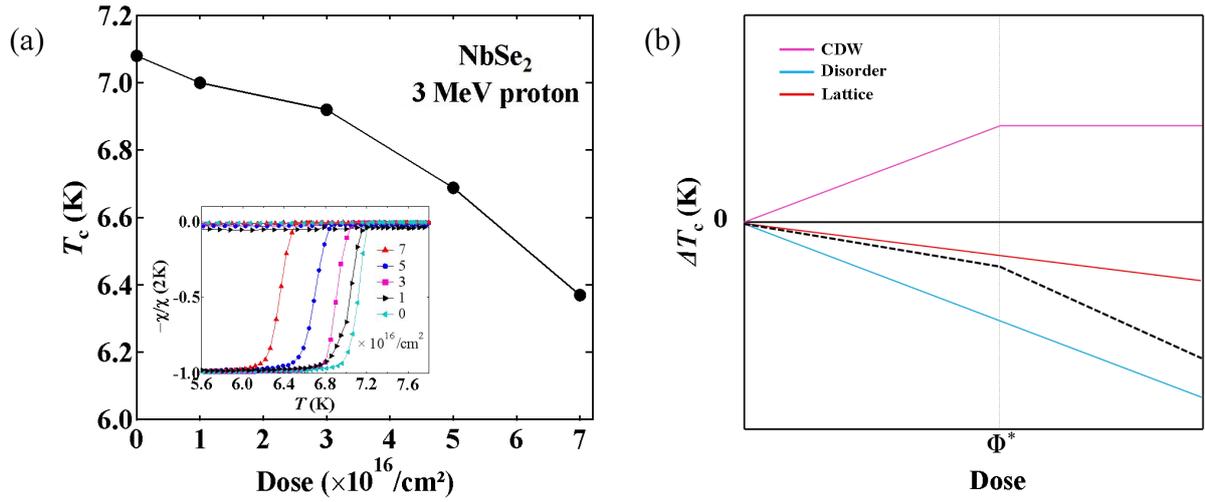

Fig. 4. (Color online) (a) The dose dependence of $T_c$ for NbSe$_2$ single crystals before and after 3 MeV proton irradiation. $T_c$ decreases monotonically with the increase in irradiation dose. The inset shows the temperature dependence of the normalized magnetic susceptibility for pristine and irradiated crystals. (b) Schematic changes of $T_c$ in NbSe$_2$ induced by CDW, disorder, and lattice expansion. $\Phi^*$ is the critical dose where CDW is completely suppressed. The combination of the three factors results in the expected change in $T_c$ (black dashed line) with increasing irradiation dose.